# Anisotropic Circular Photogalvanic Effect in Colloidal Tin Sulfide Nanosheets

Mohammad Mehdi Ramin Moayed[1,2], Fu Li[1], Philip Beck[1], Jan-Christian Schober[1], Christian Klinke[1,3,4*]

[1] Institute of Physical Chemistry, University of Hamburg, 20146 Hamburg, Germany.

[2] Deutsches Elektronen-Synchrotron DESY, Notkestrasse 85, 22607 Hamburg, Germany.

[3] Department of Chemistry, Swansea University - Singleton Park, Swansea SA2 8PP, UK.

[4] Institute of Physics, University of Rostock, Albert-Einstein-Straße 23, 18059 Rostock, Germany.

*Correspondence to: christian.klinke@uni-rostock.de

**Abstract:**

Tin sulfide promises very interesting properties such as a high optical absorption coefficient and a small band gap, while being less toxic compared to other metal chalcogenides. However, the limitations in growing atomically thin structures of tin sulfide hinder the experimental realization of these properties. Due to the flexibility of the colloidal synthesis, it is possible to synthesize very thin and at the same time large nanosheets. Electrical transport measurements show that these nanosheets can function as field-effect transistors with an on/off ratio of more than $10^5$ at low temperatures and p-type behavior. The temperature dependency of the charge transport reveals that defects in the crystal are responsible for the formation of holes as majority carriers. During illumination with circularly polarized light, these crystals generate a helicity dependent photocurrent at zero-volt bias, since their symmetry is broken by asymmetric interfaces (substrate and vacuum). Further, the observed circular photogalvanic effect shows a pronounced in-plane anisotropy, with a higher photocurrent along the armchair direction, originating from the higher absorption coefficient in this direction. Our new insights show the potential of tin sulfide for new functionalities in electronics and optoelectronics, for instance as polarization sensors.

**Introduction**

Tin sulfide (SnS) is one of the most interesting members of the IV-VI semiconductors[1, 2]. The layered orthorhombic (OR) crystal structure of this material[3, 4], which can be described as the atomic double-layered distorted rock salt structure[3, 5] with in-plane covalent bonds and vertical van-der-Waals forces[6, 7], promises many exciting properties, such as a small band gap[8, 9], mechanical stability[6, 7], and a high optical absorption coefficient[8, 10]. Furthermore, it has been observed that the optical and electrical properties of tin sulfide show an in-plane anisotropy, with a higher carrier mobility and conductivity along the zigzag direction of the crystal[1, 7, 8, 11]. Despite all of these advantages, many novel phenomena have not been experimentally explored for this material, since reducing the thickness of the crystal to atomic levels is not straightforward and the material is not easily available with a 2D geometry[2, 6, 10]. Several methods have been used so far to produce atomically thin SnS, including mechanical or chemical exfoliation of the bulk crystal[6, 11, 12], physical vapor deposition (PVD)[2, 8, 13] and atomic layer deposition (ALD)[2, 6], which resulted mainly in the production of thick and/or small flakes.

One possibility to synthesize tin sulfide is employing the colloidal approach, which shows a great degree of flexibility to tune and improve the product[14-17]. With this method, it is possible to reduce the thickness to less than 10 nm, while the lateral dimensions of such crystals are large enough for contacting them with standard lithographic methods[3, 5, 10, 18]. This allows to further investigate the novel properties of this semiconductor, such as the circular photogalvanic effect (CPGE). Based on the bulk inversion asymmetry (BIA) or the structural inversion asymmetry (SIA) of certain crystals, this effect results in a helicity dependent current at zero volt bias, when the crystal is illuminated with circularly polarized light[19, 20]. This current originates from the asymmetric population of photo-excited carriers in momentum space[20, 21]. The effect, which has been also observed in colloidal lead sulfide nanosheets, a material with comparable properties to tin sulfide[22], could be a platform for the realization of (circular) polarization sensitive photodetectors[19, 23]. It can also shine light on some fundamental properties of crystals, such as spin-orbit interactions or valley-polarization of carriers[24, 25].

Based on investigations on the colloidally synthesized SnS crystals with a quasi-2D character, we report in this letter the observation of anisotropic circular photogalvanic effects. First, the crystalline quality of these nanosheets was probed by performing field-effect measurements. By identifying the transport mechanism of the nanosheets, it was possible to improve the on/off ratio to more than $10^5$ at low temperatures. Then, circularly polarized light was employed to observe the CPGE, while the inversion symmetry of the crystal was broken by means of the substrate. Our measurements show that this effect is strongly anisotropic in the lateral dimensions with a higher CPGE current along the armchair direction. This anisotropy can be explained by a higher light absorption along the armchair direction, especially for the employed wavelength regime.

**Results**

*Introducing the material*

Colloidal SnS nanosheets with a lateral size of up to 5 µm and a thickness of less than 10 nm (~15 layers[11]) have been synthesized based on the recipe, mentioned in our previous report[18]. Figure 1a shows a transmission-electron microscope (TEM) image of the nanosheets, which were employed for our measurements. Since the nanosheets are relatively large, it was possible to lithographically place several contacts on a sheet to accurately investigate the crystal anisotropy without considering the possible differences between different nanosheets. Therefore, two pairs of gold contacts have been introduced to the nanosheets by employing electron-beam lithography. These contacts have been placed along the edges of the nanosheets, which corresponds to the zigzag (armchair) direction along the long (short) edge (Figure S1 in the Supplementary Information shows the high-resolution transmission electron microscopy HRTEM of these nanosheets which determines their crystallographic directions). The anisotropic growth of the nanosheets (faster growth along the zigzag direction) guarantees the relation between the dimensions of the sheets and their crystallographic directions[3, 18]. A schematic of such a device is shown in Figure 1b.

For the first group of measurements, two of the contacts were selected (the ones along the zigzag direction due to their higher performance). In combination with the highly doped silicon back gate, this device could be investigated as a field-effect transistor. The organic ligands on the surface of the nanosheets were not removed or replaced explicitly, in order to preserve the stability of the nanosheets, although they could marginally reduce the contacts quality.

Figure 1c shows the output characteristics of such a device at different temperatures. As the figure shows, by reducing the temperature, the current through the sheets becomes smaller.

While the conductivity (measured at 2 V bias and 0 V gate voltage) at room temperature reaches to 5 $\Omega m^{-1}$, it decreases to less than 0.005 $\Omega m^{-1}$ at 5 K. The room temperature conductivity of these nanosheets is comparable with chemically or even physically produced tin sulfide[3, 11], which confirms the high crystalline quality of the nanosheets. Figure 1d summarizes the conductivity of these sheets at different temperatures. The interesting feature, which can be observed here, is the sharp drop of the conductivity at temperatures between 100 and 150 K. This transition temperature corresponds to a change in the transport of carriers for SnS and will be discussed later in more details.

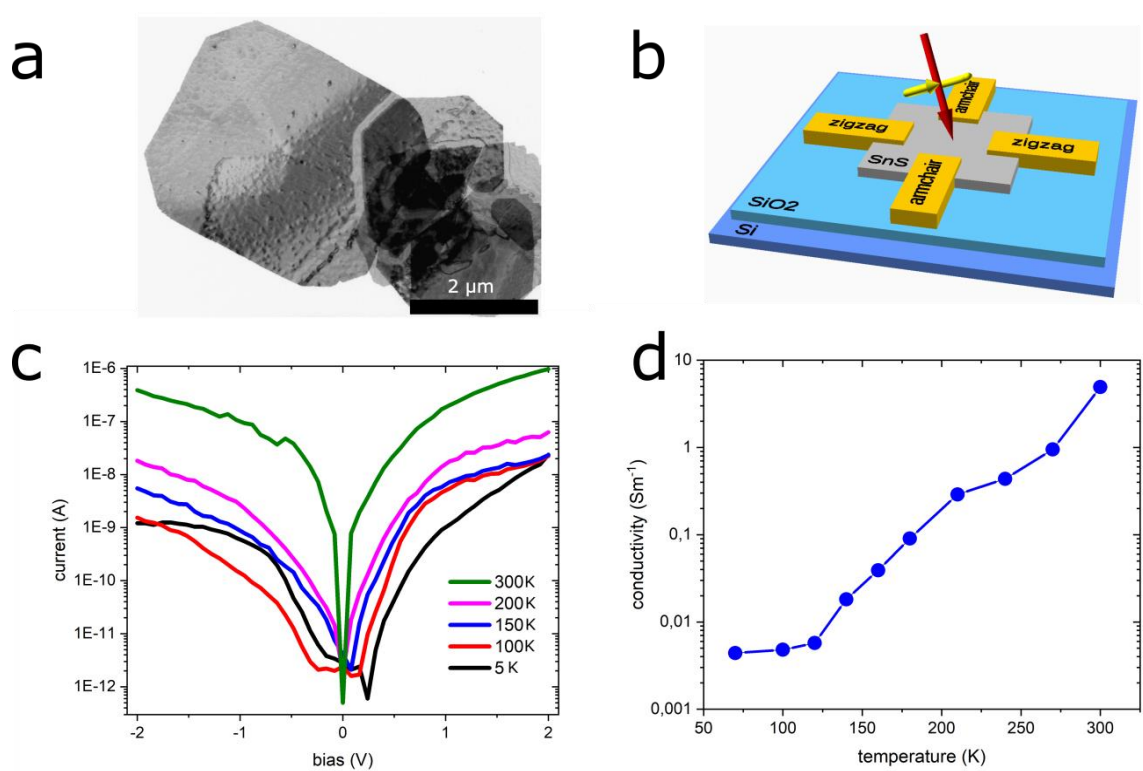

**Figure 1.** *SnS nanosheets and their electrical behavior. a) Transmission-electron microscope image of the nanosheets with lateral sizes of up to 5 µm. b) Schematic view of the device with two pairs of contacts along the zigzag and armchair directions. Field-effect measurements have been performed along the zigzag direction. c) Output characteristics of the FET at different temperatures. Here, the gate voltage is set to zero. d) Conductivity of the nanosheets at different temperatures, measured at 2 V bias and 0 V gate voltage. The conductivity decreases by reducing the temperature and becomes relatively constant below 100 K.*

Figure 2a shows the transfer characteristics of the nanosheets. First of all, it can be observed that the conductivity increases by applying negative gate voltages, which means that holes are the majority carriers of the sheets. The SnS crystal tends to form defects, acting like p-doping in the system. Because of this reason, the nanosheets represent p-type transistor behavior[3, 4, 7, 10, 11, 26]. Comparable to other forms of SnS, the on/off ratio of these nanosheets is small at room temperature, and does not exceed the value of 10 [3, 4, 10, 11, 26]. This is caused by the relatively high concentration of holes in the system, which makes the crystal highly doped and screens the gate electric field. Therefore, the devices cannot be effectively switched off[3, 11, 26]. By reducing the temperature, as can be seen in Figure 2b, the concentration of activated holes reduces and as a result, the on/off ratio increases. Below 150 K and down to 100 K, a sharp increase of the on/off ratio can be observed. This value reaches more than $10^5$, which is, to our knowledge, the highest reported on/off ratio for SnS even at low temperatures[3, 4, 10, 11, 26]. Interestingly, the increase of the on/off ratio and the decrease of the conductivity occur at a similar temperature range (100-150 K). In order to clarify the origin of these changes, the carrier concentration at different temperatures has been calculated based on the data of the conductivity and mobility ($\sigma=n\mu e$)[11], and the result is shown in Figure 2c. By fitting the values of the carrier concentration in this graph, the activation energy of the charge carriers (holes) is obtained to be around 35 meV. Based on these observations, the electrical behavior of the nanosheets can be explained. At room temperature, a high concentration of holes is available for the transport and therefore, the conductivity of the nanosheets is high. As mentioned, these holes screen the gate electric field and hinder the switching effect. By reducing the temperature below 150 K, the thermal energy of the system becomes insufficient to activate the holes, which results in a sharp decrease of the conductivity. On the other hand, the screening effect of the holes is suppressed, and the on/off ratio increases. Below 100 K, these changes are much smaller and therefore, the conductivity and the on/off ratio remain relatively constant. These results are in a very good agreement with the theoretical predictions and also the previous experimental observations based on this material[9, 11, 26, 27], confirming the very high crystal quality of our nanosheets.

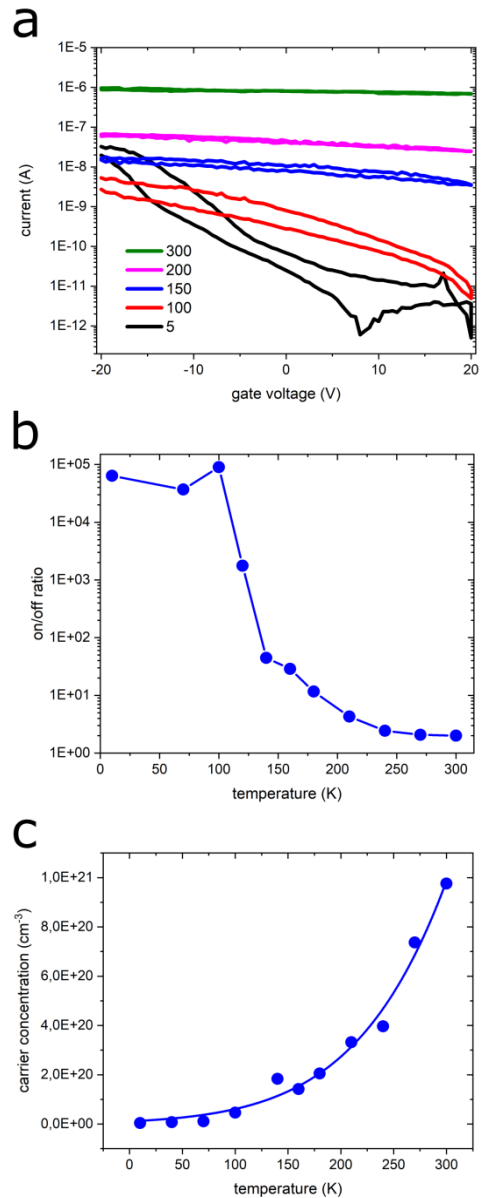

**Figure 2.** *Field-effect behavior of the nanosheets. a) Transfer characteristic of the nanosheets at different temperatures. More pronounced switching effects are observed at low temperatures. b) Temperature dependency of the nanosheets' on/off ratio. Below 100 K, the sheets can be completely switched off and the on/off ratio reaches to about $10^5$. c) Variation of the carrier concentration by changing the temperature. By increasing the temperature above 100 K, the carrier concentration starts to sharply increase.*

*Circular photogalvanic effect*

For the second group of measurements, the nanosheets were employed to investigate their response to circularly polarized light (circular photogalvanic effect). For this purpose, they were illuminated with a red laser beam ($\lambda$=630 nm), and their short circuit current was measured, while the helicity of the light was modulated with a quarter-wave plate. The beam was oriented obliquely on the nanosheets (with a small angle, limited by the setup) and the projection of the beam on the nanosheet had an azimuthal angle of 45º to the direction of the current flow (the path between the electrodes). First, the contacts along the armchair direction were selected and the photocurrent was measured along this direction. Figure 3a represents the photocurrent for different light polarizations (angle of the quarter-wave plate). As can be clearly seen, the measured photocurrent oscillates when the angle of the quarter-wave plate is changed from 0º to 360º. These oscillations represent a meaningful dependency to the light helicity. In order to understand this relation, the following equation can be employed:

$$J_{\text{total}} = J_0 + J_{\text{CPGE}} \sin(2\varphi) + J_{\text{LPGE}} \sin(2\varphi)\cos(2\varphi) \qquad (1)$$

Here, $J_0$ is the background current, which is independent of the polarization and is originating from photovoltaic effects, the Damber effect, or hot electron injection. $J_{\text{LPGE}}$ is the amplitude of the linear photogalvanic effect (LPGE), which is due to the asymmetric scattering of carriers along different conduction paths. The LPGE is a function of the quarter-wave plate angle, but it is independent of the helicity. The oscillation period of the LPGE is equal to 90°. Eventually, $J_{\text{CPGE}}$ is the amplitude of the circular photogalvanic effect (CPGE), the current which is generated as a result of the circular polarization of light and is directly proportional to the light helicity. In contrast to the LPGE, the CPGE oscillates with the period of 180° [24, 25].

Employing the above mentioned equation to fit the experimental results reveals that the measured photocurrent contains a CPGE current of 1.5 pA, which is a value comparable to similar systems like PbS nanosheets[22]. A requirement to observe the CPGE is inversion asymmetry of the crystal, which can be achieved either intrinsically by an asymmetric crystal structures (BIA) or artificially (externally) by applying a gate voltage, preparing heterostructures or mechanical strain (SIA)[19, 21, 22, 24, 25]. In our case, the *Pbmn* crystal structure of nanosheets obeys the $D_{2h}$ point group, which is inversion symmetric[1, 4, 9, 11, 12, 26]. However, different environments on two sides of the nanosheets, Si/SiO$_2$ substrate underneath and vacuum on top, cause an effective vertical asymmetry for the crystal. Since the dielectric constant of the substrate significantly differs from vacuum, distinct screening patterns are formed on different sides of the nanosheets, which modifies the band structure and breaks the inversion symmetry[21, 22, 24, 28, 29]. As a result, the symmetry of the crystal is reduced to a $C_{2v}$ one, which supports the generation of the CPGE[19, 21]. Other factors like defects or temperature gradient effects could also enhance the asymmetry of the crystal[19, 30].

The other factor, which influences the symmetry of crystals under certain conditions is the gate-electric field. For SnS, in contrast to many other materials, the gate-electric field is less effective (due to screening effects) and the asymmetric vertical interfaces play the main role to break the symmetry and to induce the CPGE[22].

Phenomenologically, the generated photocurrent can be calculated with the following equation:

$$J_{CPGE-\lambda} = \sum_{\mu} \chi_{\lambda\mu} e_{\mu} E_0^2 P_{circ} \tag{2}$$

where $\mu$ is different crystal orientations (x, y, z), $\chi$ is the CPGE second-rank pseudo tensor, which is directly affected by the crystal asymmetry, $E_0$ is the electric field (complex amplitude) of the light wave, $P_{circ}$ is the helicity of the circularly-polarized light (degree of polarization), $e = \mathbf{q}/q$ is the unit vector for the light propagation, and $q$ is the wave vector of the light in the medium[24, 31, 32]. For the $C_{2v}$ point group, the pseudo tensor $\chi$ has non-zero elements, which results in a non-zero CPGE current[21].

After detecting the CPGE along the armchair direction of the nanosheets, the experiment was repeated along the zigzag direction. These measurements were performed by employing the contacts along the zigzag direction and by keeping all the other parameters constant. As shown in Figure 3b, a significant reduction of the CPGE can be observed (from 1.5 pA along the armchair direction to 0.3 pA), which is a sign for an anisotropic CPGE.

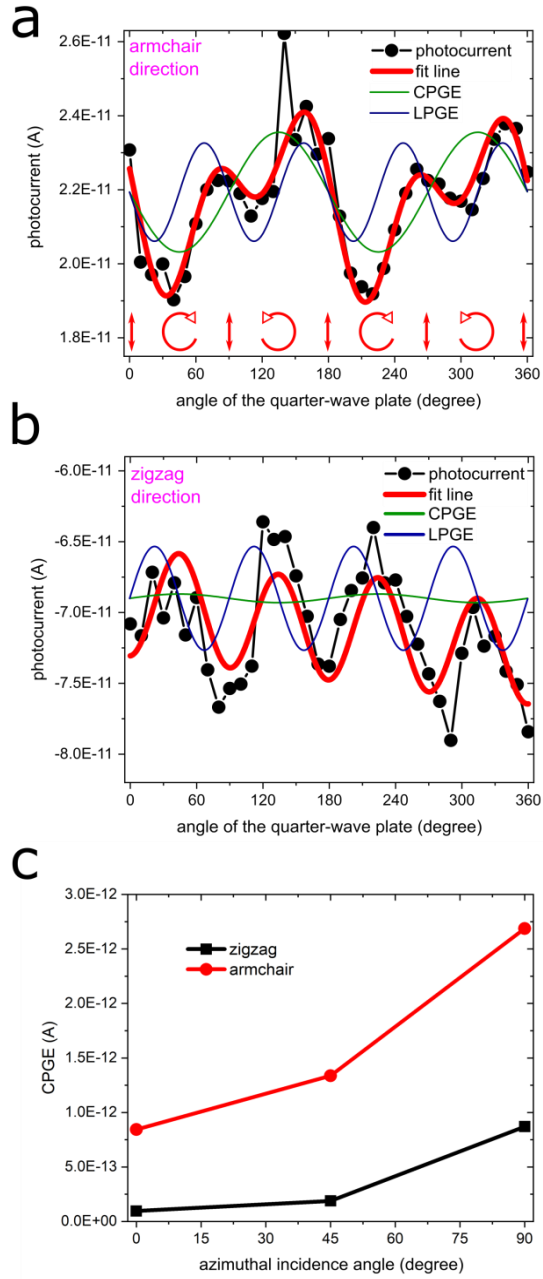

**Figure 3.** *Circular photogalvanic effect observed with SnS nanosheets. a) Variation of the photocurrent by changing the angle of the quarter-wave plate. A helicity-dependent current (CPGE) can be detected along the armchair direction (azimuthal angle of illumination: 45º). b) Variation of the photocurrent along the zigzag direction by changing the angle of the quarter-wave plate (all the other parameters unchanged). c) The detected CPGE current along different crystallographic directions and their dependence to the azimuthal incidence angle. Here, 0º (90º) corresponds to the incidence plane parallel (perpendicular) to the direction of the current in each crystallographic axis.*

To further investigate the strength of this effect in different crystallographic directions, the azimuthal angle of the incidence plane (compared to the direction of the current) was also varied. As can be seen in Figure 3c, along both directions, the CPGE current is very small when the azimuthal angle is close to zero (incidence plane parallel to the direction of the current in each crystallographic axis). It increases by increasing the azimuthal angle and becomes maximum for the azimuthal angle of 90º (incidence plane perpendicular to the current direction). In general, the CPGE current is generated transverse to the light scattering plane and perpendicular to the asymmetric direction of the crystal[31]. As a result, the azimuthal angle of the incidence affects the CPGE current according to the following equation,

$$CPGE \propto sin(\theta) \qquad (3)$$

in which, $\theta$ is the angle between the in-plane projection of the beam and the direction of the current flow[33, 34]. As can be observed, the measured CPGE currents qualitatively follow this equation, although errors of measurements or parallel effects (such as asymmetric contacts or edges) might result in quantitative deviations.

More importantly, the figure shows that along the armchair direction, the CPGE current is significantly higher than that along the zigzag direction. Such superiority can be observed for all the azimuthal angles. When the incidence plane of the laser beam is perpendicular to the direction of the current flow ($\theta=90º$), the CPGE along the armchair direction is 3.6 times higher than the CPGE along the zigzag direction.

The anisotropy of the CPGE can be explained by investigating the absorption coefficients of the crystal. As previously shown by Lin *et al.*, the absorption coefficient of SnS is higher along the armchair direction, when the crystal is excited with photon energies higher than ~1.55 eV[13]. Since the excitation energy in our experiments was around 2 eV ($\lambda$: 630 nm), it can be concluded that the conversion of the light energy to an electrical current occurs more efficiently along the armchair direction, which results in a higher CPGE current along this direction. This is in agreement with the previously reported observations about the anisotropic CPGE of black phosphorus, a material with a similar crystal structure and symmetry to SnS[19, 30].

**Conclusion**

In summary, the anisotropic circular photogalvanic effect was investigated in colloidal tin sulfide nanosheets. First, the crystal quality was probed by performing field-effect measurements. Identifying the mechanism of the defect-based p-type transport through these nanosheets led to the improvement of their on/off ratio to more than $10^5$ at low temperatures. Further, by breaking the symmetry of the crystals with asymmetric interfaces (substrate and vacuum), they showed a helicity dependent photocurrent at zero-volt bias, when they were illuminated with circularly polarized light. The observed CPGE effect was stronger along the armchair direction due to the higher absorption coefficient in this direction. Our results introduce new applications for tin sulfide in electronics and optoelectronics, such as polarization sensitive photodetectors.

**Materials and Methods**

*Preparation of 2D SnS nanosheets*

In a typical reaction, a mixture of $Sn(CH_3CO_2)_2$ (TA, 0.25 mmol), oleic acid (OA, 0.64 mmol), trioctylphosphine (TOP, 1 mmol) and octadecene (ODE, 10 mL) is added into the reaction flask, where ODE serves as solvent and OA and TOP acts as capping ligand. After continuous heating to 180°C for about 15 min, the mixture is completely dissolved. Then, the reaction solution is cooled down to 75°C, degassed and dried under vacuum for 2h. Then, the solution is heated up to 300°C and trioctylphosphine-S (TOP-S, 1M, 0.52 mL) is quickly injected into the flask. After 5 min reaction, the solution is cooled down to room temperature. The products are purified by centrifugation with toluene at 2000 rpm for 3 min (2-3 times) and then dispersed again in toluene with good dispersity for further characterization or storage.

*Device preparation*

SnS nanosheets suspended in toluene were spin-coated on silicon wafers with 300 nm thermal silicon oxide. The individual nanosheets were contacted by e-beam lithography followed by thermal evaporation of Ti/Au (1/55 nm) and lift-off.

*Transport measurements*

Immediately after device fabrication, we transferred the samples to a probe station (Lakeshore-Desert) connected to a semiconductor parameter analyzer (Agilent B1500a). All the measurements have been performed in vacuum. The vacuum chamber had a view port above the sample, which is used for sample illumination. For illumination of the nanosheets, a 20mW red laser (630 nm) with a spot size of 2mm was used. The polarization of the laser beam was controlled by a polarization filter and a quarter-wave plate.

**Acknowledgments**

M.M.R.M., and C.K. gratefully acknowledge financial support of the European Research Council via the ERC Starting Grant "2D-SYNETRA" (Seventh Framework Program FP7, Project: 304980). C.K. thanks the German Research Foundation DFG for financial support in the frame of the Heisenberg scholarship KL 1453/9-2. M.M.R.M. thanks the PIER Helmholtz Graduate School for the financial support. F.L. thanks the China Scholarship Council (CSC) also for the financial support. C.H. thanks the DFG for funding via the project SFB668 (project B17) and the North-German Supercomputing Alliance (HLRN) for computational resources.

**Conflict of Interest**

The authors declare no conflict of interest.